\documentclass[fleqn,usenatbib]{mnras}

\usepackage{newtxtext,newtxmath}
\usepackage{subcaption}
\usepackage[T1]{fontenc}

\DeclareRobustCommand{\VAN}[3]{#2}
\let\VANthebibliography\thebibliography
\def\thebibliography{\DeclareRobustCommand{\VAN}[3]{##3}\VANthebibliography}

\usepackage{graphicx}	% Including figure files
\usepackage{amsmath}	% Advanced maths commands
\newcommand{\lya}{Ly$\alpha$}

\newcommand{\kms}{$km s^{-1}$}

\newcommand{\HI}{\mbox{H\,{\sc i}}}

\newcommand{\OVI}{\mbox{O\,{\sc vi}}}

\newcommand{\CIV}{\mbox{C\,{\sc iv}}}

\newcommand{\NeVIII}{\mbox{Ne\,{\sc viii}}}

\title[\OVI\ clustering at low-$z$]{
Role of ionizing background and galactic feedback on the redshift space clustering of \OVI\ absorbers in hydrodynamical simulations
}

\author[Maitra et al.]{Soumak Maitra$^{1}$\thanks{E-mail: soumak.maitra@inaf.it},Sukanya Mallik$^{2}$\thanks{E-mail: sukanyam@iucaa.in},
	Raghunathan Srianand$^{2}$\thanks{E-mail: anand@iucaa.in},
	\\
	\\
 $^{1}$ Istituto Nazionale di Astrofisica -Osservatorio Astronomico di Trieste, Via Tiepolo 11, Trieste, Italy\\
	$^{2}$ IUCAA, Postbag 4, Ganeshkhind, Pune - 411007, India\\
}

\date{Accepted XXX. Received YYY; in original form ZZZ}

\pubyear{2015}

\begin{document}
\label{firstpage}
\pagerange{\pageref{firstpage}--\pageref{lastpage}}
\maketitle

\begin{abstract}
We explore the effect of ionizing UV background (UVB) on the redshift space clustering of low-$z$ ($z\le 0.5$) \OVI\ absorbers  using Sherwood simulations incorporating "WIND" (i.e. outflows driven by stellar feedback) only and "AGN+WIND" feedbacks. { These} simulations show a positive clustering signals up to a scale of 3 Mpc. We find that the effect of feedback is restricted to small scales (i.e $\le$2 Mpc or $\approx 200$\kms\ at $z\sim0.3$) and "WIND" only simulations produce stronger clustering signal compared to simulations incorporating "AGN+WIND" feedbacks.
{ How clustering signal is affected by the assumed UVB depends on the feedback processes assumed. For the simulations considered here} the effect of UVB is confined to even smaller scales (i.e $<$1 Mpc or $\approx 100$\kms\ at $z\sim0.3$). These scales are also affected by exclusion caused by line blending.  Therefore, our study suggests clustering at intermediate scales (i.e 1-2 Mpc for simulations considered here) together with the observed column density distribution can be used to constrain the effect of feedback in simulations. 
%while clustering at larger scales are useful for constraining the cosmological parameters.
\end{abstract}

%\begin{keywords}
%keyword1 -- keyword2 -- keyword3
%\end{keywords}
\begin{keywords}
		Cosmology: large-scale structure of Universe - Cosmology: diffuse radiation - Galaxies: intergalactic medium - Galaxies: quasars : absorption lines
\end{keywords}
%%%%%%%%%%%%%%%%%%%%%%%%%%%%%%%%%%%%%%%%%%%%%%%%%%

%%%%%%%%%%%%%%%%% BODY OF PAPER %%%%%%%%%%%%%%%%%%

\section{Introduction}

Absorption lines seen in the spectra of distant quasars allow us to probe physical conditions prevailing in the low density intervening media.
In particular, the statistics of metal line absorption can be used to constrain various feedback processes at play in the formation and evolution of galaxies. { Since the ion fractions of different elements in these absorbers are influenced by both photo- and collisional ionization, they also prove to be an important probe of the ionizing UV Background (UVB) along with the feedback processes.}
This can be done by comparing various statistics obtained from the observed spectra with those generated from the hydrodynamical simulations \citep[][]{ oppenheimer2009,tepper2011,oppenheimer2012,suresh2015,rahmati2016,Nelson2018,bradley2022, Li2022,Khaire2023}.
Recently, we \citep{Mallik2023} have shown that the frequently used statistics like column density distribution function (CDDF), distribution of b$-$parameter and fraction of \HI\ absorbers showing detectable metal absorption are not only affected by different feedback processes considered but also by the assumed ionizing UV background. As UVB in the energy range responsible for ionization of commonly detected ions like \CIV, \OVI\ , and \NeVIII (i.e 48$-$207 eV range) are observationally ill-constrained it hampers our ability to place constraints on different feedback processes. 

The next step is to study how the choice of UVB and feedback processes affects the clustering of absorbers. Considering \HI\ \lya\ forest, two-point correlation statistics (either based on transmitted flux pixels or on \HI\ absorbers) have been shown to be sensitive to the UVB \citep[for example,][]{gaikwad2017b, gaikwad2019, maitra2020b}. The dependence of two-point correlation function (2PCF) { on the $N$(\HI) threshold} for the \lya\ absorbers is mainly driven by the well-known correlation between the gas over-density and $N$(\HI). Therefore, for a given $N$(\HI) threshold changing the UVB is like changing the over-density range over which clustering is measured \citep[see][for discussions]{maitra2020,maitra2020b}. On the other hand, feedback processes are found to have little effect on the clustering of \HI\ \lya\ absorbers \citep{maitra2020b} for a given $N$(\HI) threshold.

Following \citet{Mallik2023}, here we explore the effect of feedback and UVB on the clustering of \OVI\ absorbers at $z<0.5$ using Sherwood simulations.
We consider \OVI\ because it traces the metal distribution in low density regime { and is easily} observable over a {wide redshift} range for the wavelength range typically covered by HST-COS observations. {
Note that there are several published papers in the literature that study \OVI\ absorbers in cosmological simulations \citep{cen2001,fang2001,oppenheimer2009,oppenheimer2012,rahmati2016,bradley2022,Li2022,Mallik2023}. However, none of them study redshift space clustering using absorption line components decomposed using Voigt profile fitting as done in observations\citep[see for example][]{danforth2016}.
\cite{Nelson2018} investigated 3D real space two-point correlation of \OVI, O~{\sc vii} and O~{\sc viii} ions using IllustrisTNG simulations. They did notice a flattening of clustering signals at small scales (i.e $\sim$ few 100 pkpc). This they attributed to the ion density profile being
shallower with radius compared to that of  total oxygen, total metal,
or total gas mass. While they presented a detailed investigation of how different feedback processes affect CDDF no such analysis is presented for the clustering.
In this work, we will study the more observationally motivated (i.e clustering of absorption components) 1D redshift space two-point correlation function in simulations.}
%which allows us to compare it with observational clustering studies of \OVI\ absorbers \citep[as for example in][]{danforth2016}.}
%\Soumak{Should we mention 1 sentence about why we chose to work with OVI?} 
For the simulations considered here we show that (i) the clustering of \OVI\ absorbers at 
large scales (i.e $\ge$ 2 Mpc) {is nearly} same { for all simulations examined} and not significantly influenced by the choice of UVB; 
%mainly sensitive to cosmological parameteres; %\Soumak{Is it ok to mention this since we do not show the 1st point explicitly and simply infer it?} 
(ii) In the intermediate length scales (i.e $\sim$1-2 Mpc) the clustering is influenced mostly by the assumed feedback processes and not by the assumed UVB and (iii) the strongest effect of UVB is seen mainly at smaller scale i.e $<$ 1 Mpc. 
{ While exact values of different length scale quoted above are specific for the simulations used here, this study suggests that clustering at the intermediate scales can be used as a good discriminator between models with different feedback prescriptions. }

This article is organized as follows.
In Section~\ref{Sec:Simulations}, we describe the simulation setup and the forward modelling procedure used to generate mock \OVI\ absorbers. In Section~\ref{Sec:Clustering}, we describe the estimator to compute the longitudinal (or redshift space) two-point correlation function. We study the effect of different {feedback models} and { forms of the assumed UV backgrounds} on the clustering of \OVI\ absorbers. Section~\ref{Sec:Conclusion} provides a general summary and discussions. Throughout the paper, we quote the length scales in physical units, unless mentioned otherwise. The simulations used in this work use a standard $\Lambda$CDM cosmology with cosmological parameters based on \cite{planck2014} (\{$\Omega_m$, $\Omega_b$, $\Omega_\Lambda$, $\sigma_8$, $n_s$, $h$\} = \{0.308, 0.0482, 0.692, 0.829, 0.961, 0.678\}).

\section{Simulations}\label{Sec:Simulations}
This work utilizes simulations from the Sherwood suite \citep{Bolton2017}, which were run on the smoothed particle hydrodynamics code {\sc GADGET-3}, which is a modified version of the publicly available code {\sc GADGET-2}\footnote{\url{http://wwwmpa.mpa-garching.mpg.de/gadget/}} \citep{springel2005}. {These were run within a box of size (80 $h^{-1}$ cMpc)$^3$ and having $2\times 512^3$ particles.} { The mass resolution in  this Sherwood simulation is $2.75 \times 10^8 h^{-1} M_{\odot}$(dark matter), $5.1 \times 10^7 h^{-1} M_{\odot}$(baryon) with gravitational softening length set to 1/25$^{th}$ of the mean inter-particle spacing. The initial conditions were generated on a regular grid using the N-\textsc{GEN}IC code at $z=99$ \citep[]{springel2005} using transfer functions generated by \textsc{CAMB} \citep[]{lewis2000}. The feedback models and parameters implemented in the Sherwood simulations result in a good agreement of galaxy properties (e.g., galactic stellar mass function) and redshift evolution of star formation rate density with the observations \citep[see section 3.1-3.3 in][]{Puchwein2013}. Specifically, we use two simulation feedback models: (1) "WIND" only feedback incorporating an energy-driven outflow model, and (2) "AGN+WIND" feedback model \citep[see][]{Puchwein2013}. These are initiated with the same seed and all other parameters except for the feedback models are kept the same.  While the Sherwood simulation models incorporating only wind feedback are well converged, \citet{Puchwein2013} reports that the models, including the AGN feedback, may have some issues in convergence in the mass resolution similar to the box used in this work.}
%This may not affect the results in this work as the aim of this work is to understand the effect of given feedback models and their variation on the metal absorber clustering.

{ The global metallicity of the SPH particles is obtained from the stellar mass, and we assume solar relative abundance of different elements. The Sherwood simulations were run using a spatially uniform \cite{haardt2012} UVB. However, to study the effects of different UVBs on the clustering of metal ions, we use the spatially uniform UVBs from \cite{faucher2009} (fg11) and \cite{khaire2019} (ks19q18; corresponding to a far-UV spectral index $\alpha=-1.8$) while generating the spectrum as a post-processing step.
As explained in \citet{Mallik2023} these two encompass the range of UVB predicted by various models (see their Figure 1), and the \OVI\ photoionization rate in "ks19q18" is 2.83 times higher than that in "fg11". We make an explicit assumption here that the change in gas temperature when we use different UVBs is negligible. This is a reasonable assumption as the gas in low redshift $(z \leq 0.5)$ is dominated by the adiabatic expansion cooling and not the photoheating.

The volume, mass resolution, star formation, and metal production prescriptions and feedback models are broadly consistent with most of the simulations that are used to study metal absorbers in IGM \citep[for example,][]{oppenheimer2009, oppenheimer2012, tepper2011, tepper2013, Nelson2018}. Different statistics like column density distribution, Doppler parameter and system-width distribution of \OVI\ absorbers extracted from this simulation agree well with the observation for the "AGN+WIND" model using "fg11" UVB \citep[see Figure 7-9 in][]{Mallik2023}. Therefore, these simulations are adequate for exploring the metal absorber clustering as the focus of this paper. We use simulation snapshots at $z=0.1, 0.3$, and 0.5 for this work. }

\subsection{Generating mock spectra with \OVI\ absorbers}

\begin{table*}
    \centering
 %   \begin{center}
   \caption{ Number of O{\sc vi} components per unit redshift interval ($dn/dz$) for different O{\sc vi} column density thresholds $N$(\OVI)$>N_{min}$(in cm$^{-2}$). The values have been quoted for AGN+Wind (A+W) and Wind only (W) feedback models and also for ks19q18 and fg11 UV backgrounds at different redshifts. We also quote the $dn/dz$ value for the combined sample of O{\sc vi} absorbers at different redshifts used in this work.}
     \begin{tabular}{c|cc|cc|cc|cc}
     \hline
log $N_{min}$  &    \multicolumn{2}{c|}{z=0.1} & \multicolumn{2}{c|}{z=0.3} & \multicolumn{2}{c|}{z=0.5} & \multicolumn{2}{c|}{Combined}\\

 & A+W & W & A+W & W & A+W & W & A+W & W \\
%parameters &  Sherwood  &  & Sherwood  &  &    Massive Black & \\
%to         &   only &  & "AGN+WIND"  &  &    "AGN+WIND"      & \\
%compare    &  feedback  &  & feedback  &  &    feedback      & \\
   \hline
    \hline
%    & &UVB -> & ks18q18 & & &\\
\multicolumn{9}{c}{\underline{For "ks19q18" UVB}}\\
13.0 & 5.68 & 2.51  & 9.25 & 3.47 & 13.82 & 5.13 & 9.89 & 3.80\\
13.5 & 2.77 & 1.59  & 4.46 & 2.11 & 7.19 & 3.19 & 4.97 & 2.36 \\
14.0 & 0.25 & 0.54  & 0.58 & 0.65 & 1.20 & 0.99 & 0.71 & 0.74 \\
\\
\multicolumn{9}{c}{\underline{For "fg11" UVB}}\\
13.0 & 13.58 & 2.48 & 20.31 & 3.03 & 25.39 & 4.40 & 20.20 & 3.38 \\
13.5 & 6.46 & 1.58 & 10.04 & 1.78 & 13.35 & 2.57 & 10.21 & 2.01 \\
14.0 &  0.62 & 0.54 & 1.16  & 0.59 &  2.07 & 0.82 & 1.34 & 0.66 \\

    \hline
    \end{tabular}
 %   \end{center}
  
    \label{tab:dn_dz_OVI}
\end{table*}

Here, we provide a general overview of our procedure for generating mock \OVI\ absorption spectra  \citep[details can be found {in}][]{Mallik2023}. We shoot spatially random sightlines across the simulation box, to generate the \OVI\ absorbers. The 1D field { quantities} like hydrogen density ($n_H$), temperature ($T$), and peculiar velocity ($v$) are computed along these sightlines {(in equispaced grids of width $\sim$ 7kms$^{-1}$)} using SPH smoothing of the nearby particle values \citep{monaghan1992}. The \OVI\ number density $n_{\rm OVI}$ for these SPH particles is computed from the particle $n_H$, $T$,  $Z$ and given UVB using cloudy \citep[version 17.02 of the code developed by][]{ferland1998} assuming optically thin conditions. { The ionization fractions have been computed assuming both photo-ionization and collisional ionization equilibrium.} The 1D $n_{\rm OVI}$ field is then estimated along the sightline by SPH smoothing the particle information. We use the $n_{\rm OVI}$ along with the $T$ and $v$ fields to generate the \OVI\ absorption spectra along the sightlines. We then forward model these simulated spectra by convolving them with a Gaussian profile of FWHM=17 $km s^{-1}$ 
({ as that of HST-COS}) to simulate instrumental smoothing and then adding a Gaussian noise corresponding to SNR=30 per pixel. We generate 20,000 such sightlines for {each case}, as we vary the UVB and feedback models. While generating the \OVI\ absorption spectra, we do not simulate the doublets or mimic contamination effects from other absorbing species for the sake of simplicity.

For the clustering analysis, we decompose the \OVI\ absorption into distinct components using the automated Voigt profile fitting routine {\sc viper} \citep[see][for details]{gaikwad2017b}. Note that {\sc viper} was written with the purpose of statistical analysis of the \HI\ \lya\ forest and as of now, it does not support the simultaneous fitting of multiple transitions from a single ion. So we consider only the strongest transition line while performing the Voigt profile decomposition, thus raising the caveat of possible underestimation of component structures associated with highly saturated transitions.  We consider only absorption components that are detected at a rigorous significant level \citep[RSL, as described in][]{keeney2012} more than 4. For the SNR considered here this translates to a 90\%  completeness for $N$(\OVI)$>10^{13}$ cm$^{-2}$ \citep[see figure 6 of][]{Mallik2023}.

In order to measure the two-point correlation function of absorbers, it is important to measure the mean number of absorbers per unit redshift interval (i.e. dn/dz). We compute this by finding the total number of \OVI\ absorbers above a certain column density threshold in the simulated spectra and then dividing it by the total redshift path length covered by all the spectra.
The values of dn/dz for our simulations at three different redshifts and two different UVBs are summarized in Table~\ref{tab:dn_dz_OVI}.  For all the column density thresholds (log$N_{min}=13, 13.5$ and 14), the obtained dn/dz for \OVI\ is systematically higher for AGN+WIND simulations when we use "fg11" UVB. However, the effect of UVB is not that significant when we consider the WIND only simulations \citep[see][for detailed discussions]{Mallik2023}.
%Discussions on why this happens can be found in \citet{Mallik2023}.

\section{Clustering of \OVI\ absorbers}\label{Sec:Clustering}

%In this section, we will calculate the longitudinal 2PCF profile for the \OVI\ absorbers and investigate its dependence on ionizing background and feedback models. 

\subsection{Longitudinal 2PCF}\label{Sec:2PCF}

We follow a procedure similar to { that} described in \cite{maitra2020b} (for \lya\ absorbers) to compute the longitudinal 2PCF for the simulated \OVI\ absorbers. The 2PCF is expressed as a probability excess of finding an \OVI\ pair with respect to a random distribution of the absorbers at a certain longitudinal physical separation $r_{\parallel}$. It is calculated using the estimator,
\begin{equation}
		\xi_{\parallel}( r_{\parallel})=\frac{<DD>}{<RR>}-1 \ ,
	\end{equation}	 
where "DD" and "RR" are the data-data and random-random pair counts of \OVI\ absorbers at a separation of $r_{\parallel}$ \citep[see][]{kerscher2000}.  The total data-data pairs $DD$ at each $r_{\parallel}$ bin are summed over 20,000 sightlines for each model  and then normalized with $n_D (n_D-1)/2$ (where $n_D$ is the total number of absorbers). 
%To generate random sightlines, the \OVI\ absorbers are distributed randomly along the sightlines. 
The number of random \OVI\ absorbers to be distributed along each sightline is drawn from a random Poisson distribution about the mean number of absorbers obtained from dn/dz in Table~\ref{tab:dn_dz_OVI}. The RR is computed using 1000 such random sightlines for each data sightline to minimize the variance. All the random pair counts are also normalized with the total number of pair combinations ($n_R (n_R-1)/2$). 
%\begin{subfigure}{0.195\textwidth}
%    \centering\includegraphics[width=\textwidth]{plots/slit_configs/slit_J0127-0550_41.pdf}
     %\caption{Caption text 1}
%  \end{subfigure}
\begin{figure*}
\centering
    \begin{subfigure}{0.45\textwidth}
    \includegraphics[viewport=0 40 320 255,width=7cm, clip=true]{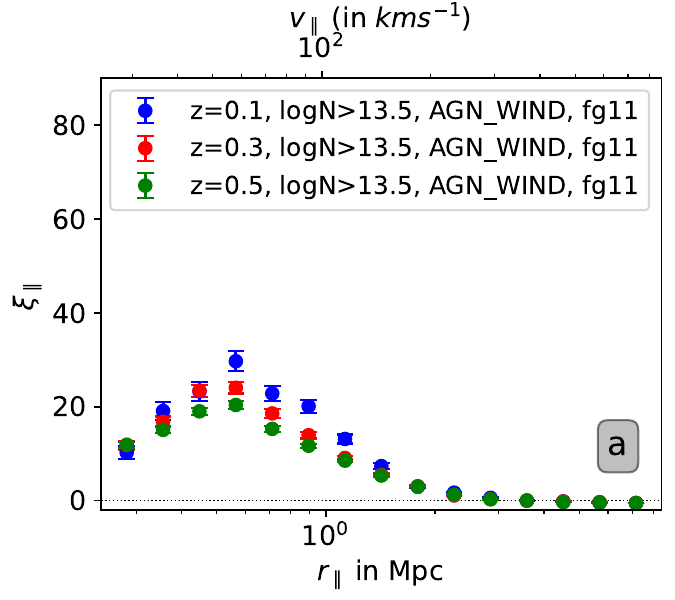}
    \end{subfigure}
   \begin{subfigure}{0.45\textwidth} \includegraphics[viewport=0 40 320 255,width=7cm, clip=true]{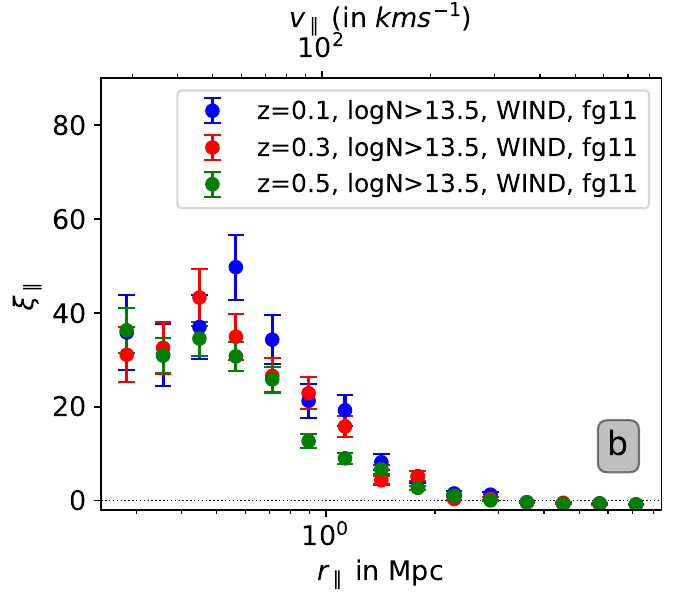}%
    \end{subfigure}

\begin{subfigure}{0.45\textwidth}
\includegraphics[viewport=0 5 320 255,width=7cm, clip=true]{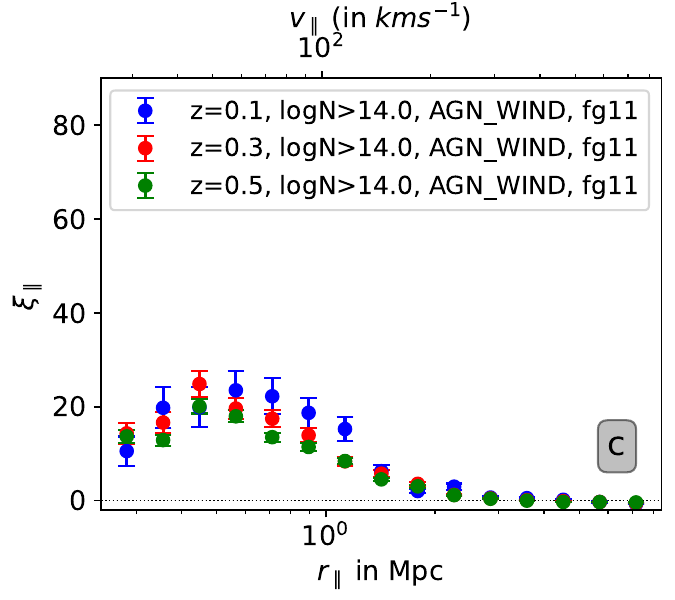}%
    \end{subfigure}
 \begin{subfigure}{0.45\textwidth}   \includegraphics[viewport=0 5 320 255,width=7cm, clip=true]{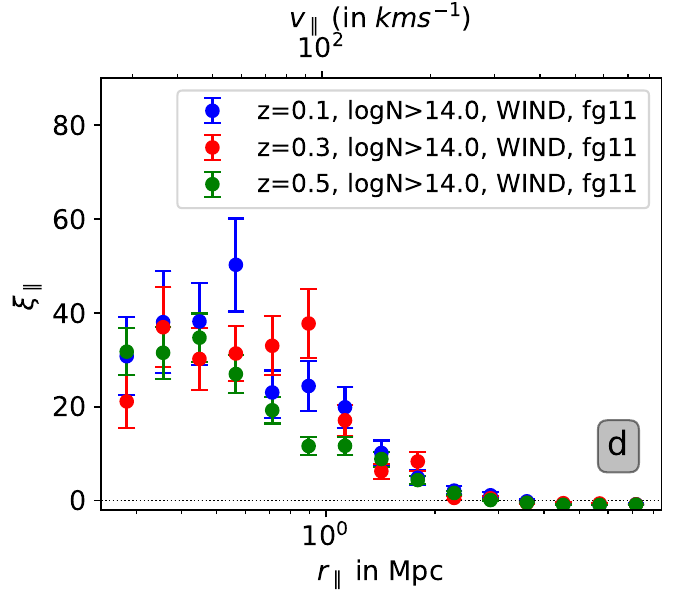}%
    \end{subfigure}
    \caption{ Longitudinal 2PCF profiles (in physical coordinates) for \OVI\ computed at redshifts $z=0.1,\ 0.3$ and 0.5. The plots have been shown for the "fg11" photo-ionizing UV background. The left and right panels show the plots for AGN+Wind and Wind-only feedback models, respectively. The top and bottom panels show the 2PCF profile for \OVI\ absorbers having column densities $N$(O~{\sc vi})> $10^{13.5}$ and $10^{14}$ cm$^{-2}$, respectively.  }%The plots vary over column density thresholds (between plots in the top and bottom panels). 
    %Note that OVI detection has been done for SNR=30 combining 20000 sightlines for each case.}
    \label{2PCF_redshifts_physical}
\end{figure*}

In Fig.~\ref{2PCF_redshifts_physical}, we plot the longitudinal 2PCF of \OVI\ absorbers at $z=0.1,0.3$, and 0.5 as a function of physical separation for 2 different feedback models (WIND and AGN+WIND) and $N$(O~{\sc vi}) thresholds ($10^{13.5}$ and $10^{14}$ cm$^{-2}$). The errorbars in the longitudinal 2PCF is the larger of the two errors: one-sided Poissonian uncertainty corresponding to $\pm 1\sigma$ or the bootstrapping error for all the data-data pairs. { We detect a positive 2PCF up to a length scale of 3 pMpc. The 2PCF profile decreases monotonically with scale for $r_\parallel>500$kpc. At smaller scales ($r_\parallel<500$kpc), we observed a suppression in the 2PCF. Such a suppression is not seen in the clustering profile of metal ions studied by \citet{Nelson2018}.
Such a suppression, we see here, is also seen in the longitudinal clustering of \lya\ absorbers \citep[refer to Fig.~3 in][] {maitra2020} while using individual absorption components. We ascribed it to the finite width of thermally broadened (along with instrumental broadening) \OVI\ absorbers that prevent multicomponent Voigt profile decomposition below certain velocity (and hence physical distance) scales.
%
%which sets a lower limit on the scale of identification of multiple components in \OVI\ absorptions. 
It is seen that the AGN+WIND model typically produces wider \OVI\ absorbers in comparison to wind-only models \citep[check Fig.~9 of][for the comparison between the width of \OVI\ absorbers between the different feedback models]{Mallik2023}{}{}, which might explain the stronger small-scale suppression seen in the case of AGN+WIND model.}

When we consider the simulation with AGN+WIND feedback we see the following trends: 
(i)  We detect clear clustering signal up to 3 Mpc ($\approx 300$\kms\ for $z\sim0.3$).
(ii) There is no clear redshift evolution in the $\xi(r_\parallel>1.5)$ Mpc. (iii) We see a strong redshift evolution ($\xi$ increasing with decreasing $z$) at small scales. While the same trend is seen for both $N$(O~{\sc vi}) thresholds used, the errorbars associated with $\xi(r_\parallel$) are larger when we consider higher column densities. (iv) Also, for a given scale we do not see a significant trend of $\xi$ increasing with the $N$(O~{\sc vi}) threshold. When we compare the two simulations the amplitude of $\xi({r_\parallel})$ is higher in the case of WIND only simulations for small scales. However, WIND only simulations do not show a clear trend with redshift for $\xi(r_\parallel)$.

For the study involving the effect of UVB and feedback processes on clustering, we decide to combine 20,000 sightlines each from the three $z=0.1,0.3$, and 0.5 simulation snapshots for calculating the clustering amplitudes. The random sightlines are generated using the dn/dz from the combined sample (see Table~\ref{tab:dn_dz_OVI}). Our decision to combine the three redshift sightlines allow us to create a more realistic sample of mock spectra covering \OVI\ absorbers in the redshift range $0.1\leq z\leq 0.5$ \citep[see][]{danforth2016, maitra2020b}. It is, however, important to note that while generating these sightlines, we do consider the redshift evolution of the UVB across $z=0.1,0.3,0.5$ and not the same UVB across the entire redshift range.

%The first noticeable thing in the 2PCF profile is the suppression of clustering signal at scales $<500$pkpc, irrespective of the redshift, feedback model or $N_{OVI}$. At larger scales, we find positive clustering signals up to a length scale of $2.5$Mpc. We also don't find a significant evolution of the 2PCF profile with redshift, except for the AGN+WIND case for $N_{OVI}>10^{13.5}$cm$^{-2}$ absorbers. Another interesting thing to note is the fact that the 2PCF profile doesn't depend significantly on the column density thresholds. We also find that the clustering signals are seemingly stronger at lower redshifts. We still decide to combine different redshift sightlines in this study to simulate a more realistic depiction of the redshift ranges covered by \OVI\ absorbers in observations. 

\begin{figure*}

    \includegraphics[viewport=0 40 320 300,width=6cm, clip=true]{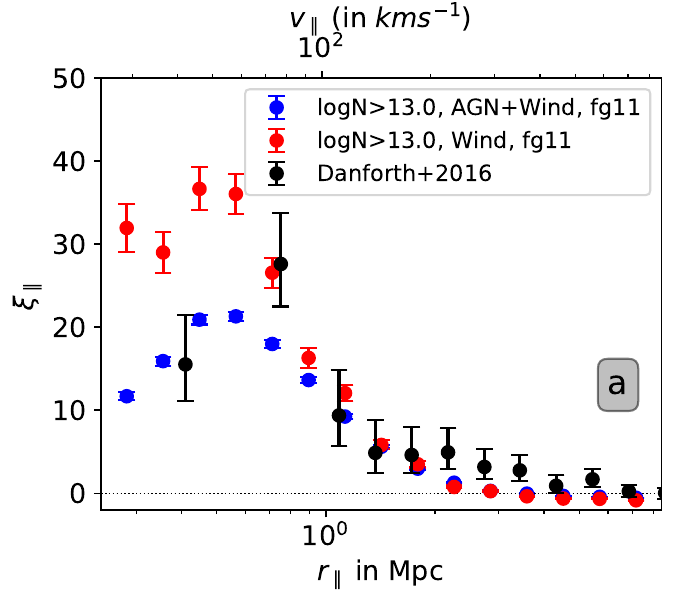}%
    \includegraphics[viewport=0 40 320 300,width=6cm, clip=true]{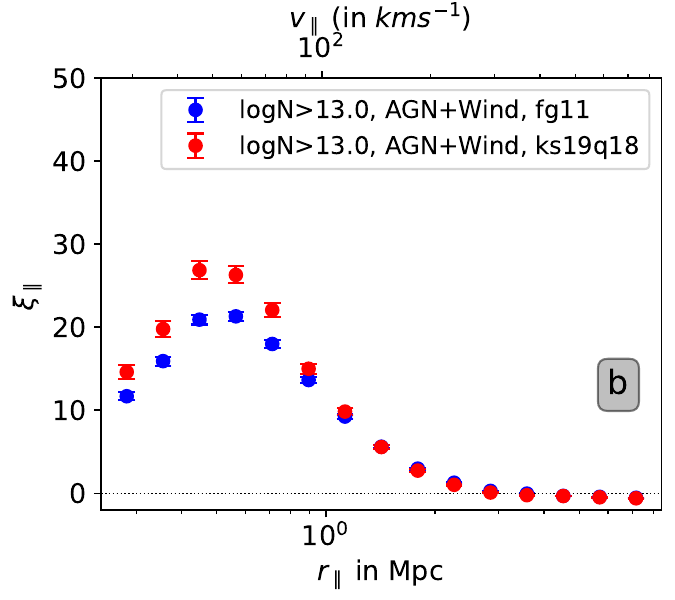}%
    \includegraphics[viewport=0 40 320 300,width=6cm, clip=true]{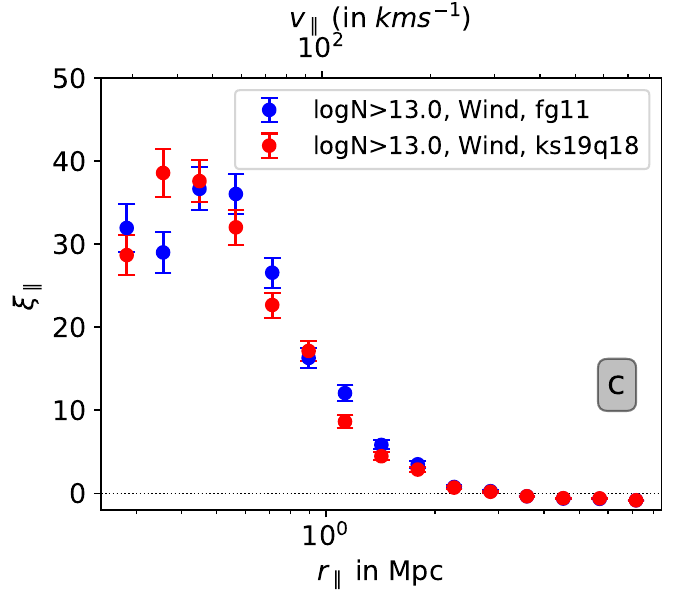} 
    
    \includegraphics[viewport=0 40 320 260,width=6cm, clip=true]{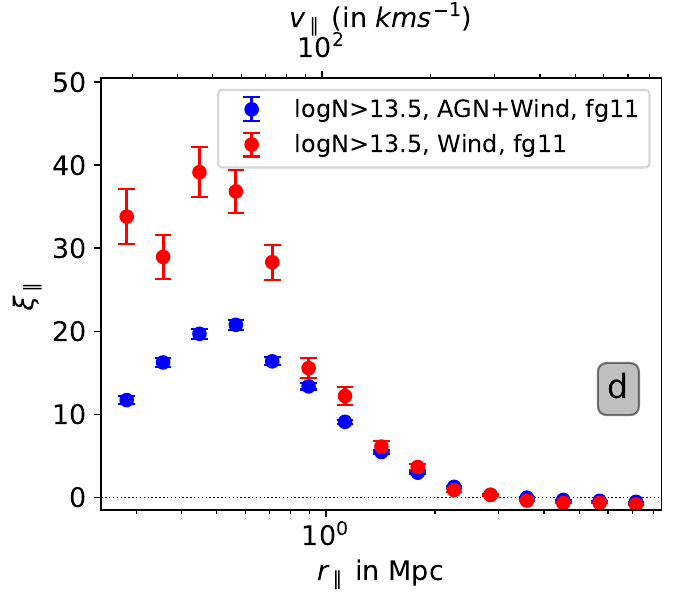}%
    \includegraphics[viewport=0 40 320 260,width=6cm, clip=true]{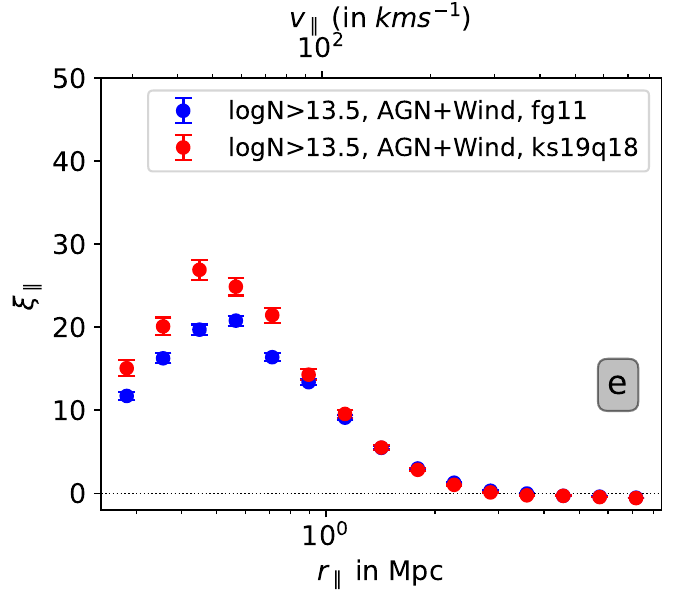}%
    \includegraphics[viewport=0 40 320 260,width=6cm, clip=true]{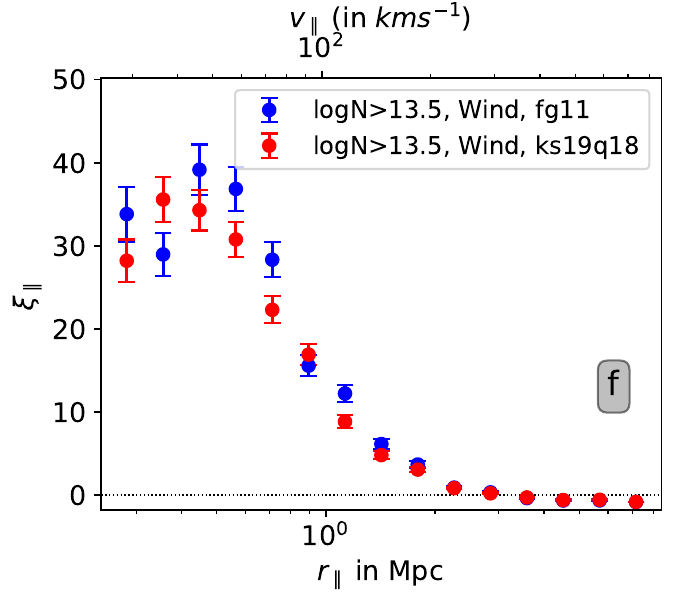} 

    \includegraphics[viewport=0 0 320 260,width=6cm, clip=true]{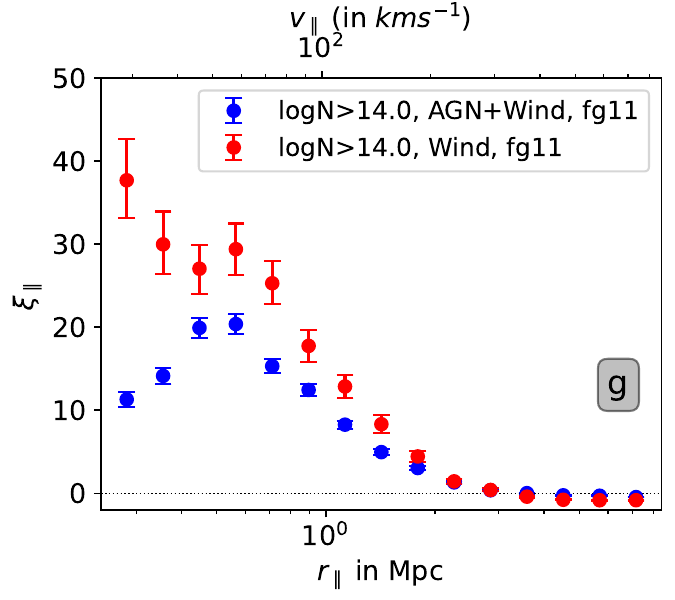}%
    \includegraphics[viewport=0 0 320 260,width=6cm, clip=true]{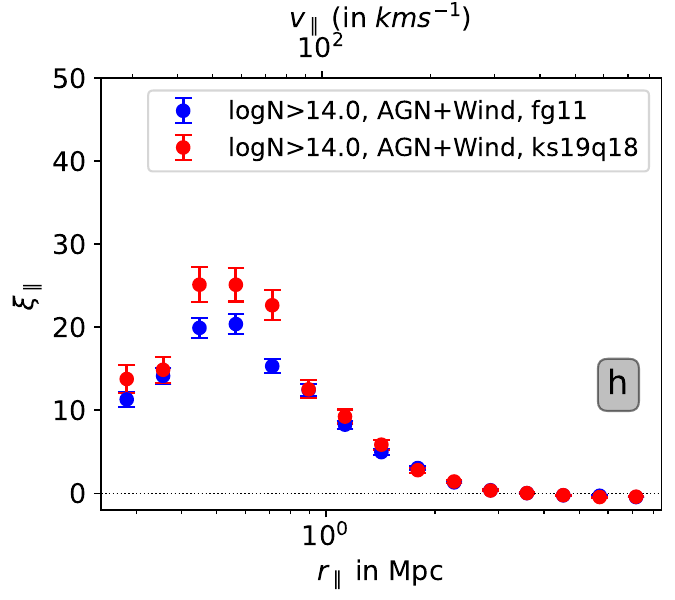}%
    \includegraphics[viewport=0 0 320 260,width=6cm, clip=true]{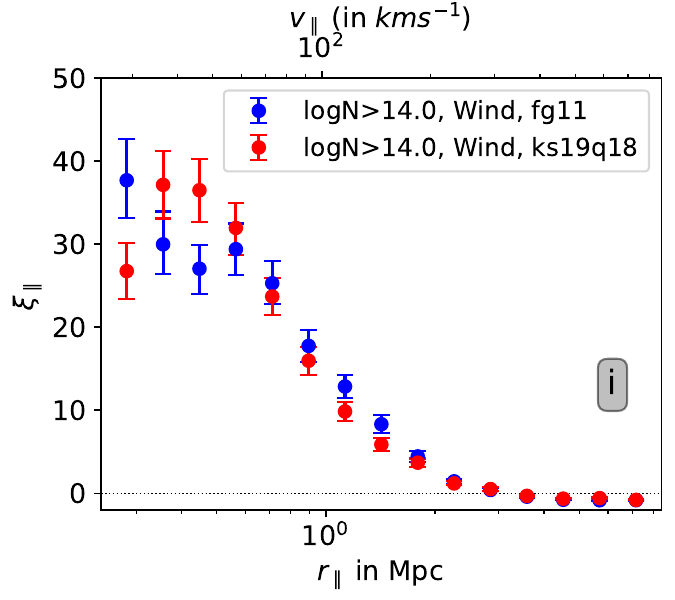}     
   
    \caption{ Longitudinal 2PCF profiles for \OVI\ (in physical units; along with velocity scale corresponding to a mean redshift of 0.3). The left panels show the comparison between the two feedback models AGN+Wind and Wind-only. On the middle and right panels, we show the comparison between the "fg11" and "ks19q18" UV backgrounds for AGN+Wind (middle panels) and Wind-only (right panels) feedback models. The top, middle, and bottom panels show the 2PCF profile for \OVI\ absorbers having column densities $N$(O~{\sc vi})> $10^{13}$, $10^{13.5}$, and $10^{14}$ cm$^{-2}$, respectively. %Note that OVI detection has been done for SNR=30 combining 20000 sightlines each from z=0.1, 0.3 and 0.5 simulation snapshots.
    In panel (a), we also plot the longitudinal clustering of \OVI\ absorbers from observations using HST-COS spectra \citep{danforth2016}. 
    %The lowest column density detected in the observed data sample is $\approx 10^{13}$cm$^{-2}$.
    }
    \label{2PCF_physical}
\end{figure*}

\subsection{Effect due to feedback}\label{Sec:Feedback}

Panels (a), (d), and (g) in Fig.~\ref{2PCF_physical} compare the longitudinal 2PCF profiles between WIND and AGN+WIND feedback models. Irrespective of the %column density thresholds of the absorber 
$N_{min}$ used, we find the clustering amplitudes to be greater in the case of WIND feedback model up to a scale of 2 Mpc. 
{
It is also evident from this figure that this result is independent of \OVI\ column density threshold assumed. 
From the detailed descriptions presented in Section 2.2 of \citet{Mallik2023}, we note that  for the "Diffuse"\footnote{Please refer to section 2.2 of \citet{Mallik2023} for the definition of different phases.} and "WHIM" phases the fraction of SPH particle with \OVI\ ion fraction in excess of 0.01 and their mean metallicities are higher in the case of AGN+WIND models compared to "WIND" only models.  On the other hand, the same quantities are higher in the ``Condensed" 
and "Hot halo" phases for the WIND only model.
\citet{Mallik2023} have concluded that, in the simulations considered here, we expect detectable \OVI\ absorption also from both ``Diffuse" and ``WHIM" phases in the case of ‘WIND + AGN’ Sherwood simulations. On the other hand, \OVI\ absorption will predominantly detected from the ``Hot-halo" or ``Condensed" phase, in the ‘WIND’ only simulation.  Such a picture is also consistent with low dn/dz seen in the case of WIND only simulations (see Table~\ref{tab:dn_dz_OVI}).
}
%
%One cannot ascribe this behaviour to the { greater} abundance of high column density absorbers in the case of WIND model, since we already saw that \OVI\ clustering amplitudes are relatively independent of column density thresholds. One of the reasons might be the difference in physical sources of the absorbers between WIND and AGN+WIND feedback models. While the \OVI\ absorbers originate usually from the WHIM or diffuse phase in the AGN+WIND feedback model, in case of WIND only feedback, they originate predominantly from hot halo or condensed regions ({ check Sec.~2.2 on the phase space distribution of \OVI\ in \cite{Mallik2023} for a detailed description}). 
Since the "Hot halo" and "Condensed" phases typically trace denser and more clustered density fields, this might explain the higher clustering amplitudes seen in the case of WIND only feedback model. 

For the sake of completeness, we also compare the simulated clustering amplitudes with that  observed for the HST-COS sample \citep[][]{danforth2016} in panel (a). It is to be noted that the sample of \OVI\ absorbers used to compute the observed clustering is not based on column density thresholds. The weakest absorber detected in the sample has $N_{\rm OVI}\simeq 10^{13}$cm$^{-2}$ \citep[see Fig.10 of][]{danforth2016}. We compare this observed 2PCF profile with the simulated 2PCF profile with absorbers having $N_{\rm OVI}>10^{13}$cm$^{-2}$. Within the errorbars, the observed \OVI\ clustering amplitude matches well with simulations up to a length scale of 2 Mpc (matches better for WIND model at $\approx 0.7$ Mpc and with AGN+WIND model at $\approx 0.4$ Mpc). We find an excess of positive clustering at scales up to 6 Mpc in the observations as opposed to $\simeq$2 Mpc in the case of simulations. 
%This may suggest that the simulations miss the large scale power in the clustering of \OVI.
This either suggests {an under-prediction of the volume filling} of metals or a deficiency in the large-scale power if \OVI\ absorbers predominantly originate from the circumgalactic medium of galaxies in the simulations. { Also we use constant SNR in our simulations. For proper comparison with observations, we need to consider SNR distribution and wavelength-dependent absorption line detection sensitivity for each observed spectrum \citep[As in the case of][for \lya\ absorbers]{maitra2020b}. Such an exercise is not needed for the present study as we are mainly interested in probing the difference in the clustering properties of the two simulations.}

It is also evident from the figure that beyond 2 Mpc the two simulations produce nearly identical 2PCFs { for all  three \OVI\ column density thresholds considered}. As the cosmological parameters and initial conditions are same for both the simulations, it is { most likely} that beyond 2 Mpc  the clustering is dominated by the cosmological parameters {for the feedback models considered in our simulations}.

\subsection{Effect due to photoionizing backgrounds}\label{Sec:UVB}

%\citet{maitra2020b} have shown that 
In the case of \lya\ forest it has been found that small scale clustering for a given $N$(\HI)\ threshold depends on the assumed UVB \citep{maitra2020, maitra2020b}. In particular, it has been found that the clustering is stronger when the ionization rate is higher, as in this case, for a given $N$(\HI)\ one probes higher density range. 
Here, we explore whether such a dependence  is clearly present in the case of \OVI\ absorbers as well.
In panels (b), (e), and (h) we plot the 2PCF profiles for two different UVBs for the AGN+WIND feedback models for three $N$(\OVI) thresholds. 

{ As can be seen from Table~1 of \citet{Mallik2023}, the \OVI\ photoionization rate is 2.8 times higher for the "ks19q18" UVB compared to that of "fg11".  How this affects the number of absorbers and individual \OVI\ column density measurements depend on the physical state of the gas (i.e. temperature and density).  It is clear from Table~\ref{tab:dn_dz_OVI}
that in the case of "AGN+WIND" models there is a substantial increase in the number of absorbers when we use "fg11" background. The lower photoionization rate typically increases the \OVI\ ion fraction this leads to the enhancement in the column density and number of weak absorption components above the detection threshold.
However, the effect of UVB on the number of detected absorbers is less significant in the case of WIND only models. This is because most of the \OVI\ absorbers detected in this case originate from high density regions where collisional excitation has a stronger influence.  We once again refer the readers to section 2 of \citet{Mallik2023} for details.
}

%
%Comparing the \OVI\ 2PCF profiles in panels b,c, e and f, we find that a slight UVB dependence is seen when AGN+Wind feedback model is considered.
%
It is clear that 
the effect of UVB is seen only for $r_\parallel\le0.8$ Mpc for simulations considered here. 
In this range, the 2PCF amplitudes are, in general, slightly larger in the case of "ks19q18" background. 
Even though the amplitudes of 2PCF is large in the case of WIND only feedback models the errors are also larger. Within the measurement errors 
the UVB dependence even at small scales is not that significant compared to what we see in the case of AGN+WIND feedback models. This behavior is, again, commensurate with the finding of \citet{Mallik2023} that CDDF of \OVI\ depends weakly on UVB for the WIND only feedback models while it shows a strong dependence in the case of simulations with AGN+WIND feedback (also see Table~\ref{tab:dn_dz_OVI}). 
It is also evident that the difference in the 2PCF at small scale found between two feedback models for a given UVB is much larger than the scatter introduced by changing UVB for a given feedback model. This again indicates that the clustering at small scales may more strongly depend on the feedback model ({ specifically on what is the temperature, density,  and metallicity of the regions where \OVI\ absorption originate}) than on the assumed UVB.
%Thus it appears that by fitting CDDF and 2PCF one will be able to place better constraints on the feedback models and cosmological parameters compared to what one would have done using only CDDF.

%\textcolor{blue}{Sukanya: 

It is evident from Table~\ref{tab:dn_dz_OVI} that there are more \OVI\ systems and Voigt profile components identified when we use "fg11" UVB in the case of AGN+WIND models. 
 Significant difference in the two point correlation is seen only over  $r_\parallel<1$ Mpc. To understand this we explore the simulation results for $z\sim0.3$ in some detail. Out of the 20000 sightlines considered we identify 5765 components when we use "ks19q18" UVB and 12662 components for "fg11" UVB with 3446 components being common between the two UVB.
%
%Although the number of components increases to more than double in "fg11" UVB, the clustering is affected in velocities $<200$ km s$^{-1}$ because the newly found absorbers in "fg11" arise either near the existing absorbers in "ks19q18", or in the sightlines where no absorption was detected for "ks19q18" UVB. For example, at z=0.3, "ks19q18" UVB has 5765 components, whereas the number of components in "fg11" is 12662; among them, only 3446 are common components between the two UVB. Most of these components reside in multi-component systems, with a median value of 2 components per system. The width of these common systems increases (median value of system-width increases by $16\%$ at z=0.3) in "fg11" UVB because of absorption arising from broader 
%regions.
When we use the "fg11" UVB, $\sim75\%$ 
of the newly detected components  are found to be within 2 Mpc of the common absorbers. Around 10\% are found close to the common absorbers but with separations more than this. The remaining new absorbers are distributed along sightlines that did not show \OVI\ absorption when we use "ks19q18" UVB. Most of the newly detected \OVI\ systems also require multiple Voigt profile components. This explains why the effect of UVB is restricted to small
scales. 
%
%when we use "fg11" UVB are also found near (within 200 km s$^{-1}$) the common absorbing systems and around $10\%$ of them found beyond 200 km s$^{-1}$ of any of the common absorbers, resulting in a minor contribution of UVB variation in large scale clustering. The rest of the new absorbers arises in the sightlines where absorption was not detected in "ks19q18" UVB. This set of absorbers is also mostly multi-component. Still, the system width is lower than (around $6\%$) the system width of the common absorbing systems, therefore contributing to the clustering in the lower length scale. 
This together with the lack of such a trend in "WIND" only simulations indicates that the exact range over which the effect of UVB can be seen may depend on the metal distribution in the simulations.
%considered. %\Soumak{Should we switch to Mpc here instead of km/s? It should be better to use the same units throughout.}
%}
%fact that \OVI\ absorbers predominantly trace diffuse or WHIM phases and the absorbers in the diffuse phase are more sensitive to a change in photoionizing backgrounds, as compared to other phases.

\section{Discussions}\label{Sec:Conclusion}

The observed column density distribution has been frequently used to constrain different feedback processes in hydrodynamical simulations. Recently it has been shown that the simulated CDDF is also influenced by the assumed UVB. Thus, by only using the observed CDDF one will not be able to lift the degeneracy between the effect of feedback and UVB used in the simulations. 

Considering hydrodynamical simulations with two different feedback processes and two different UVBs, we show that observed clustering of \OVI\ at different scales can be used to constrain the photo-ionizing background, feedback processes or the cosmological parameters. In the models considered here, the effect of UVB is restricted to very small scales ($<$1 Mpc) compared to the scale that are influenced by feedback processes (1-2 Mpc). Our results suggest that combining CDDF with two-point correlation function, one will be able to place better constraints on the feedback processes and the UVB.

{ While arriving at these results we have not incorporated effect of the ionizing radiation originating from local sources, in particular for \OVI\ absorption originating from the CGM. \citet{upton-sanderbeck18} have shown that local stellar high energy radiation from  Milky Way-like galaxies influence the ionization 
state of the gas up to 10-100 kpc. \citet{oppenheimer2018} have explored the influence of flickering AGNs on the ionization state of metals in the CGM. They suggested, under plausible models, this could increase the \OVI\ column densities around star-forming galaxies out to 150 kpc. Thus it appears that inclusion of the local ionizing radiation fields in the star forming regions could not have affected the above results.}

{
The length-scale up to which a feedback model can influence the 2PCF may also depend on the strength of the feedback assumed in the model. Recently SIMBA simulations including bi-polar jet-mode feedback have shown widespread disturbances can be introduced. Such feedback models not only heat the matter over large scales but also can spread metals to lower density regions.  \citet{bradley2022} have shown the introduction of such a feedback naturally reproduces the observed \OVI\ CDDF (in particular at the high column density end). As discussed in \citet{Mallik2023}
unlike other simulations, in SIMBA simulations with jet-mode feedback, a high fraction of gas (i.e $\sim70\%$ compared to $\sim$30\% found in other simulations) is present in the WHIM phase from which most of the high ions originate. Our study indicates that the  2PCF of \OVI\ obtained for the AGN+WIND model that spreads metals and influences the physical state of gas over larger scale, shows reduced amplitude compared to that of the WIND only models. 
 Therefore, we expect the 2PCF predicted by the SIMBA simulations run with jet feedback is expected to be different compared to that from other simulations.
Therefore, simultaneous comparison of CDDF and 2PCF of models with the corresponding observed distributions is important to gain much better insight on the feedback processes.
}

The clear segregation in scales over which different effects influence the inferred two-point correlation function { seen for the two simulations used here} need to be further explored using models with a wider range of feedback prescriptions. In particular, it will be interesting to establish the existence of different scales over which UVB and feedback processes influence the two-point correlation using the whole range of feedback models considered by \citet{Nelson2018} for studying the large dispersion they produce in CDDF.  

It is also known that how effective the UVB in controlling the ionization fraction of a given ion depends on the gas temperature. { For example, if \OVI\ absorption originate predominantly from the collisionally  ionized gas then we expect the dependence of the UVB to be minimal.}
Sherwood simulations are known to produce higher temperatures for the \OVI\ absorbers \citep[see][for discussions on this]{Mallik2023tur}. The above mentioned exercise using TNG simulations will allow us to address this issue as well.
%\begin{itemize}
%    \item Positive clustering up to scale 3Mpc, independent of feedback or UVB. Misses larger scale power up to 6Mpc seen in observation.
%    \item Effect of feedback 2Mpc, effect of UVB 1Mpc. Exclusion $\leq$ 0.5Mpc.
%    \item Clustering is independent of column density.
%\end{itemize}

{ Lastly, we notice that the observed 2PCF for \OVI\ absorbers has a positive signals up to $\sim$6 pMpc. However, simulations considered here have positive signal only up to $\sim$3 pMpc. It will be important to understand the origin of this  difference. As mentioned above correlation analysis performed using a wide range of simulation suites available in the IllustrisTNG simulations will be very useful in this regard.}

\section*{Acknowledgements}

SM acknowledges support from PRIN INAF - NewIGM programme. We acknowledge the use of HPC facilities PERSEUS and PEGASUS at IUCAA. We would like to extend our gratitude to Kandaswamy Subramanian for useful discussions and Prakash Gaikwad for providing the automated Voigt profile fitting code {\sc viper}. 
We would also like to mention that the Sherwood simulations were performed using the Curie supercomputer at the Tre Grand Centre de Calcul (TGCC), and the DiRAC Data Analytic system at the
University of Cambridge, operated by the University of Cambridge High
Performance Computing Service on behalf of the STFC DiRAC HPC Facility
(www.dirac.ac.uk). This was funded by BIS National E-infrastructure
capital grant (ST/K001590/1), STFC capital grants ST/H008861/1 and
ST/H00887X/1, and STFC DiRAC Operations grant ST/K00333X/1. DiRAC is part of
the National E-Infrastructure.

%%%%%%%%%%%%%%%%%%%%%%%%%%%%%%%%%%%%%%%%%%%%%%%%%%
\section*{Data Availability}

The observed \OVI\ clustering is taken from \citet{danforth2016} and the
HST-COS data products used to compute this clustering can be accessed from https://archive.stsci.edu/prepds/igm/.



\bibliographystyle{mnras}
\bibliography{main} % if your bibtex file is called example.bib

%\begin{thebibliography}{99}
%\end{thebibliography}

%%%%%%%%%%%%%%%%%%%%%%%%%%%%%%%%%%%%%%%%%%%%%%%%%%

%%%%%%%%%%%%%%%%% APPENDICES %%%%%%%%%%%%%%%%%%%%%

%\appendix



%%%%%%%%%%%%%%%%%%%%%%%%%%%%%%%%%%%%%%%%%%%%%%%%%%

% Don't change these lines
\bsp	% typesetting comment
\label{lastpage}
\end{document}